# STABILITY OF BOUNDARY LAYER FLOW BASED ON ENERGY GRADIENT THEORY

HUA-SHU DOU, WENQIAN XU

*Faculty of Mechanical Engineering and Automation, Zhejiang Sci-Tech University,*
*Hangzhou, Zhejiang Province 310018, China*
*huashudou@yahoo.com; yelanxun@126.com*

BOO CHEONG KHOO

*Department of Mechanical Engineering, National University of Singapore,*
*Singapore 119260*
*mpekbc@nus.edu.sg*

The flow of the laminar boundary layer on a flat plate is studied with simulation of Navier-Stokes equations. The mechanisms of flow instability at external edge of the boundary layer and near the wall are analyzed using the energy gradient theory. The simulation results show that there is an overshoot on the velocity profile at the external edge of the boundary layer. At this overshoot, the energy gradient function is very large which results in instability according to the energy gradient theory. It is found that the transverse gradient of the total mechanical energy is responsible for the instability at the external edge of the boundary layer, which induces the entrainment of external flow into the boundary layer. Within the boundary layer, there is a maximum of the energy gradient function near the wall, which leads to intensive flow instability near the wall and contributes to the generation of turbulence.

## 1. Introduction

In the past half century, much progress has been achieved in the understanding on the physics of boundary layer flow with the development of numerical methodologies and experimental techniques [1-4]. However, further work on this problem are still needed for the purpose of theoretical study and engineering applications.

Dou and co-authors have proposed an energy gradient theory to analyze the boundary layer flow with Blasius boundary layer profile, and they found that the gradient of the total mechanical energy becomes very large toward the wall, which is responsible for the sustenance of turbulence in boundary layer flow [5-8]. Sambasivam and Durst simulated the incompressible fluid over a flat plate with full Navier-Stokes equations and "free parallel flow" external boundary condition [9]. It is found that there is an overshoot on the velocity profile, the velocity at the edge of the boundary layer is larger than the specified free stream velocity. They suggested that the observed overshoot in the velocity profile is expected to play a significant role in determining the stability of the flow and the dynamics of transition to turbulence.

In this study, the boundary layer flow over a flat plate is simulated. The energy gradient theory is used to analyze the flow instability in the said boundary layer flow. The physics of intense (flow) instability near the wall and the entrainment of fluid at external edge are studied.

## 2. Revisiting the Energy Gradient Theory

Dou and co-authors have proposed an energy gradient theory to study the mechanisms of flow instability and turbulent transition [5-8]. In this theory, the relative magnitude of the total mechanical energy of fluid particles gained and the energy loss due to viscous friction in a disturbance cycle determines the disturbance amplification or decay, i.e., whether instability occurs or not. For a given flow, a stability criterion is expressed as,

$$K \frac{v'_m}{V} < \text{Const}, \qquad K = \frac{V \frac{\partial E}{\partial n}}{V \frac{\partial E}{\partial s} + \varphi},$$

where K is the energy gradient function, which is a local dimensionless parameter and equivalent to a local Reynolds number. It is also the ratio between the transverse gradient of the total mechanical energy and the work done by shear stresses, for the volume of fluid flowing past unit area of cross section along the streamline direction. $E = p + 0.5\rho V^2$ is the total mechanical energy where V is the total velocity, p is the static pressure, $\rho$ is the density. $v'_m$ is the amplitude of the disturbance of velocity, $\varphi$ is energy dissipation function, and n and s are along the transverse and streamwise direction, respectively.

## 3. Results and Discussions

### 3.1. *Streamwise velocity profiles*

The laminar flow in a boundary layer is simulated employing the full Navier-Stokes equation. The incoming flow on a flat plate and the external flow of the boundary layer are set to be uniform. The wall is set as no-slip and the outflow is Neumann boundary condition.

The streamwise velocity profiles at a local Reynolds number of $10^4$ are shown in Fig. 1. An enlarged view is shown in Fig.2 in which the coordinates are resized. It can be found from these pictures that there is an overshoot near the external edge of the boundary layer, which is similar to the result of Sambasivam and Durst [9]. The overshoot of velocity profile is about 1.2% higher than the incoming free stream velocity. Although this overshoot is small, it may play an important role in the mechanism of flow instability of boundary layer.

In the external flow far from the boundary layer, the velocity is almost the same as the inlet velocity. Within the boundary layer, the velocity near the wall is much smaller than the inlet velocity due to flow retardation. A vertical flow is formed up to the external edge of the boundary layer. The accumulated fluid is located at the external edge, and thus an overshoot is induced, which shows that the velocity is slightly larger than the external flow velocity.

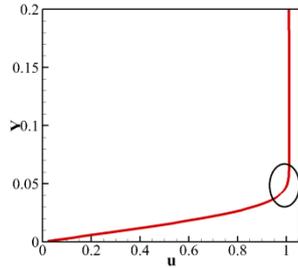 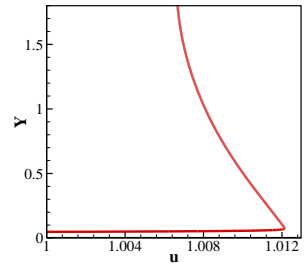

Fig. 1. Streamwise velocity profiles at a local Reynolds number $Re_x=10^4$.

Fig. 2. The zoomed-in streamwise velocity profiles of Fig. 1.

### 3.2. *Discussion on energy gradient function*

The mechanical energy gradient along the transeverse direction of streamline ($\partial E/\partial n$), the mechanical energy gradient along the tangential direction of streamline ($\partial E/\partial s$), and the velocity gradient along the transverse direction of streamline ($\partial U/\partial n$) at different local Reynolds number $Re_x=3000$, 6000, and 10000 are shown in Fig. 3-4. It can be seen from Fig. 3(a) that since the fluid inside the boundary layer obtains energy from the external fluid, the $\partial E/\partial n$ is always positive and has its maximum within the boundary layer. With the increase of the local Reynolds number, the $\partial E/\partial n$ near the wall decreases, and its maximum moves gradually away from the wall. As shown in Fig 3(b), $\partial E/\partial s$ is always negative within the boundary layer due to the effect of viscous dissipation. The velocity gradient near the wall decreases, whereas the velocity gradient in the transverse direction is as shown in Fig. 3(c). It can thus be seen that the velocity gradient near the wall decreases with the Reynolds number.

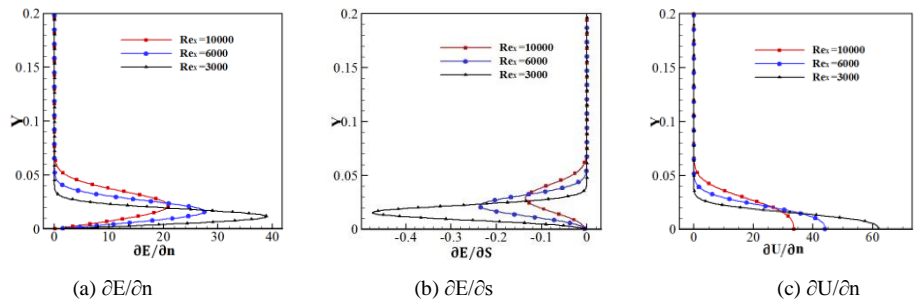

(a) $\partial E/\partial n$      (b) $\partial E/\partial s$      (c) $\partial U/\partial n$

Fig. 3. Parameter distributions at different Reynolds number: $Re_x=3000$, $Re_x=6000$, $Re_x=10000$.

The distribution of K along the transverse direction of the boundary layer is shown in Fig. 4. Within the boundary layer, the $K_{max}$ is induced near the position where the transverse energy gradient is greatest. Since the value of K is very large at this location, the capacity to amplify a disturbance is strong. It plays an important role in leading to instability and transition to turbulence in the boundary layer. As the local Reynolds number increases from 3000 to 10000, the maximum of K increases from 3000 to nearly 10,000. Thus, as the Reynolds number increases, the flow will become more unstable.

It can also be observed from Fig.4 that a local peak of K exists at the external edge of the boundary layer, which corresponds to the velocity distribution as shown in Figs.1 and 2. Since the location of $K_{max}$ can set to amplify a disturbance, the $K_{max}$ near the overshoot is responsible for the entrainment of the external flow into the boundary layer.

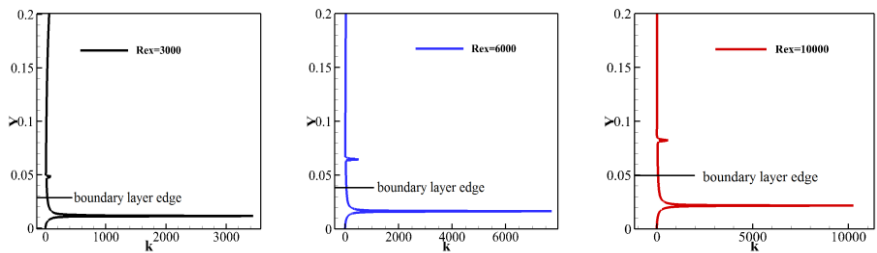

Fig. 4. The K value at different Reynolds number $Re_x$=3000, $Re_x$=6000, and $Re_x$=10000.

## 4. Conclusions

1. There is a very small overshoot of the velocity profile at the external edge of the boundary layer. Near the overshoot, there is a peak in the energy gradient function. It is believed that this peak of K plays an important role in determining the fluid entrainment from external flow into the boundary layer.

2. Within the boundary layer near the wall, a maximum of K is induced, which has the capacity to amplify a disturbance. Thus, it plays important role in generating turbulence in the boundary layer flow.

3. With the increase of the local Reynolds number $Re_x$, the value of K increases. Thus, at higher Reynolds number, the flow is more unstable at the downstream.